\documentclass[aps,prl,reprint,amssymb,showpacs,superscriptaddress,notitlepage,twocolumns]{revtex4-1}
\usepackage{bm}
\usepackage{graphicx}
\usepackage{dcolumn}
\usepackage{braket}
\usepackage{natbib}
\usepackage{amsmath}
\usepackage{color}
\usepackage{ulem}
\usepackage{siunitx}
\usepackage{mathtools,amssymb}

\begin{document}
\title{Ultrafast, low-energy, all-optical switch in polariton waveguides}
\author{D. G. Su\'arez-Forero}
\email{Present address: Joint Quantum Institute, University of Maryland, College Park, MD, USA}
\affiliation{CNR NANOTEC, Institute of Nanotechnology, Via Monteroni, 73100 Lecce, Italy}
\affiliation{Dipartimento di Fisica, Universit\`{a} del Salento, Strada Provinciale
Lecce-Monteroni, Campus Ecotekne, Lecce 73100, Italy}
\author{F. Riminucci}
\affiliation{Dipartimento di Fisica, Universit\`{a} del Salento, Strada Provinciale
Lecce-Monteroni, Campus Ecotekne, Lecce 73100, Italy}
\affiliation{Molecular Foundry, Lawrence Berkeley National Laboratory, One Cyclotron Road, Berkeley, California, 94720, USA}
\author{V. Ardizzone}
\email{vincenzo.ardizzone@nanotec.cnr.it}
\affiliation{CNR NANOTEC, Institute of Nanotechnology, Via Monteroni, 73100 Lecce, Italy}
\author{A. Gianfrate}
\affiliation{CNR NANOTEC, Institute of Nanotechnology, Via Monteroni, 73100 Lecce, Italy}
\author{F. Todisco}
\affiliation{CNR NANOTEC, Institute of Nanotechnology, Via Monteroni, 73100 Lecce, Italy}
\author{M. De Giorgi}
\affiliation{CNR NANOTEC, Institute of Nanotechnology, Via Monteroni, 73100 Lecce, Italy}
\author{D. Ballarini}
\affiliation{CNR NANOTEC, Institute of Nanotechnology, Via Monteroni, 73100 Lecce, Italy}
\author{G. Gigli}
\affiliation{CNR NANOTEC, Institute of Nanotechnology, Via Monteroni, 73100 Lecce, Italy}
\author{K. Baldwin}
\affiliation{PRISM, Princeton Institute for the Science and Technology of Materials, Princeton Unviversity, Princeton, NJ 08540}
\author{L. Pfeiffer}
\affiliation{PRISM, Princeton Institute for the Science and Technology of Materials, Princeton Unviversity, Princeton, NJ 08540}
\author{D. Sanvitto}
\affiliation{CNR NANOTEC, Institute of Nanotechnology, Via Monteroni, 73100 Lecce, Italy}
\affiliation{INFN, Sez. Lecce, Via per Monteroni, Lecce, Italy}
\begin{abstract}
    The requirement for optical-electrical-optical conversion of signals in optical technologies is often one of the majors bottleneck in terms of speed and energy consumption. The use of dressed photons (also called polaritons), that allows for intrinsic sizable interactions, could significantly improve the performances of optical integrated elements such as switches or optical gates. In this work we demonstrate the ultrafast switch of a laser coupled into a polaritonic waveguide triggered by an optical pulse resonant with the same dispersion but at a lower energy. Our experiments show two effects capable to interrupt the transmission of the laser in two different time ranges: a sub-picosecond time range due to the optical Stark effect, and a picosecond range governed by the creation of a charge reservoir. In the latter regime we found that at certain power of excitation the activation of dark states allows for a long persistence of the switching much beyond the bright exciton lifetime.
\end{abstract}

\maketitle
A fundamental component in the development of optical data processing is the ultrafast all-optical switch. Indeed, there is a growing interest in the development of fully optical devices as promising alternative for the latency of electro-optical systems, especially for applications in which electronics is only needed for amplification or modification of optical processes. Prototypes of systems capable to switch an optical signal in ultra-short times have been demonstrated in microring resonators\cite{Sethi2014}, photonic crystals\cite{Nozaki2010}, III-V semiconductor quantum wells\cite{Takahashi2000,Li2007}, graphene\cite{Li2014}, combined systems of gold and graphene\cite{ono2020} and exciton-polaritons systems \cite{Ma2020}. Exciton-polaritons, or polaritons, are hybrid light-matter quasi-particles arising when an electromagnetic mode confined in an optical resonator is coupled to an excitonic transition. Polaritons have demonstrated great potentiality for the development of all-optical technologies such as transistors\cite{Leyder2007,Ballarini2013}, routers\cite{Marsault2015}, all-optical switches\cite{Ardizzone2013} and polarization rotators\cite{Gnusov2021}. Among all the platforms sustaining polaritons, semiconductor waveguides\cite{Walker2013} (WG) could be the cornerstone for building integrated polaritonic circuits. They are attracting for their capability to sustain fast-propagating modes with high quality factors, and importantly, with simple fabrication processes. For example, optical confinement is achieved through total internal reflection with no need for growing thick distributed Bragg reflectors (DBRs) as it is the case for standard semiconductor microcavities. Waveguide geometry also allows the manipulation of the polariton-polariton interaction (and hence the optical nonlinearity) through external electric fields\cite{Rosenberg2016,Rosenberg2018,Suarez-Forero2021}. Integrated devices such as polariton lasers\cite{Jamadi2018,Suarez-Forero2020-2} have also been demonstrated using polariton waveguides. Here we demonstrate an ultrafast switching off of a polariton flow by means of another optical pulse that uses the same excitation path.\\

As it is well known any excitonic system undergoes the so called optical Stark effect when hit by a strong optical pulse off resonance with the semiconductor optical transition. In case of polaritons this effect results in an abrupt loss of the strong coupling due to the change in the exciton resonance\cite{Hayat2012,Lange2018}. As a consequence the polariton mode shifts in energy, regaining its original position only after the pulse is gone. In this work we will exploit this effect to turn off a cw laser for as long as the pulse duration. As we will see, a second effect can also take part in this process, depending on the pulse power. While for low power the Stark effect dominates the dynamics, for higher power an exciton reservoir is created due to two photons absorption. This adds a longer recovery time caused by the blueshift induced by the exciton-exciton interactions and similarly observed in Ref. \cite{Giorgi2012}. At the highest powers not only a significant reservoir is created, but also several dark excitonic states which lead to very long recovery times beyond our laser repetition rate. Such dark states have also been discussed in other works \cite{Walker}. 

\begin{figure*}[!htb]
    \centering
    \includegraphics[width=1.9\columnwidth]{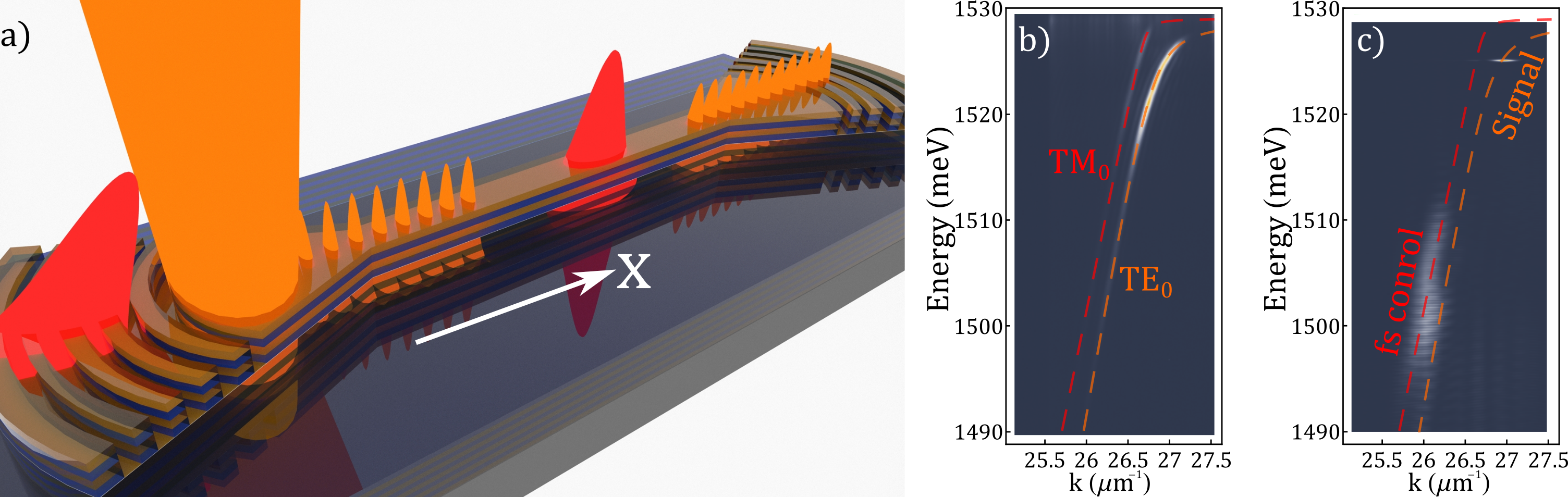}
    \caption{a) Sketch of the switch-off mechanism. A CW laser (orange) is injected resonantly inside the WG, creating a continuum flux of polaritons. When the fs pulse (red) couples into the system the transmission is interrupted via two different effects: the Optical Stark effect that modulates the Rabi splitting of the polariton mode, and the creation of a charge-carriers reservoir via two-photons absorption that quenches the transmission of the CW by drastically shifting the exciton energy. b) Angular and energy resolved photoluminescence emission of the system collected from the outcoupling grating. The theoretical fits of the $TE_0$ and $TM_0$ modes are indicated by dashed orange and red lines, respectively. c) Far field emission of the outcoupling grating when both the CW and fs lasers are resonantly injected into the system.}
    \label{sketch}
\end{figure*}

We use a WG with 12 GaAs quantum wells (QWs) 20 nm thick, separated by Al$_{0.4}$Ga$_{0.6}$As barriers of 20 nm, covered with a 10 nm GaAs capping layer. The structure has a 500 nm cladding of Al$_{0.8}$Ga$_{0.2}$As. To reduce the mode volume, a microwire structure 40 nm deep, 1 $\mu$m wide and 140 $\mu$m long is etched on the surface. A pair of focusing gratings is fabricated at both ends of the microwire for the injection and extraction of the signal and control beams. A sketch of the experimental realization is displayed in Fig.~\ref{sketch}a and a detailed description of the full experimental setup can be found in the SM. The polaritonic flux inside the WG is interrupted when a laser pulse (control) at lower energy is coupled inside the structure. Figures \ref{sketch}b-c show the guided modes dispersion and the energies and in-plane momenta at which the signal and control lasers are coupled into the system. As it can be observed in the figure, the signal laser is tuned to inject population in the $TE_o$ mode at 1525.14 meV, 4.9 meV below the excitonic resonance. The control laser is injected into the $TM_o$ mode and tuned at around $1500$ meV; far from the exciton resonance to avoid the presence of population created by photoluminescence and to exploit the Optical Stark effect (Fig.~\ref{sketch}c). \\

\begin{figure}
    \centering
    \includegraphics[width=0.9\columnwidth]{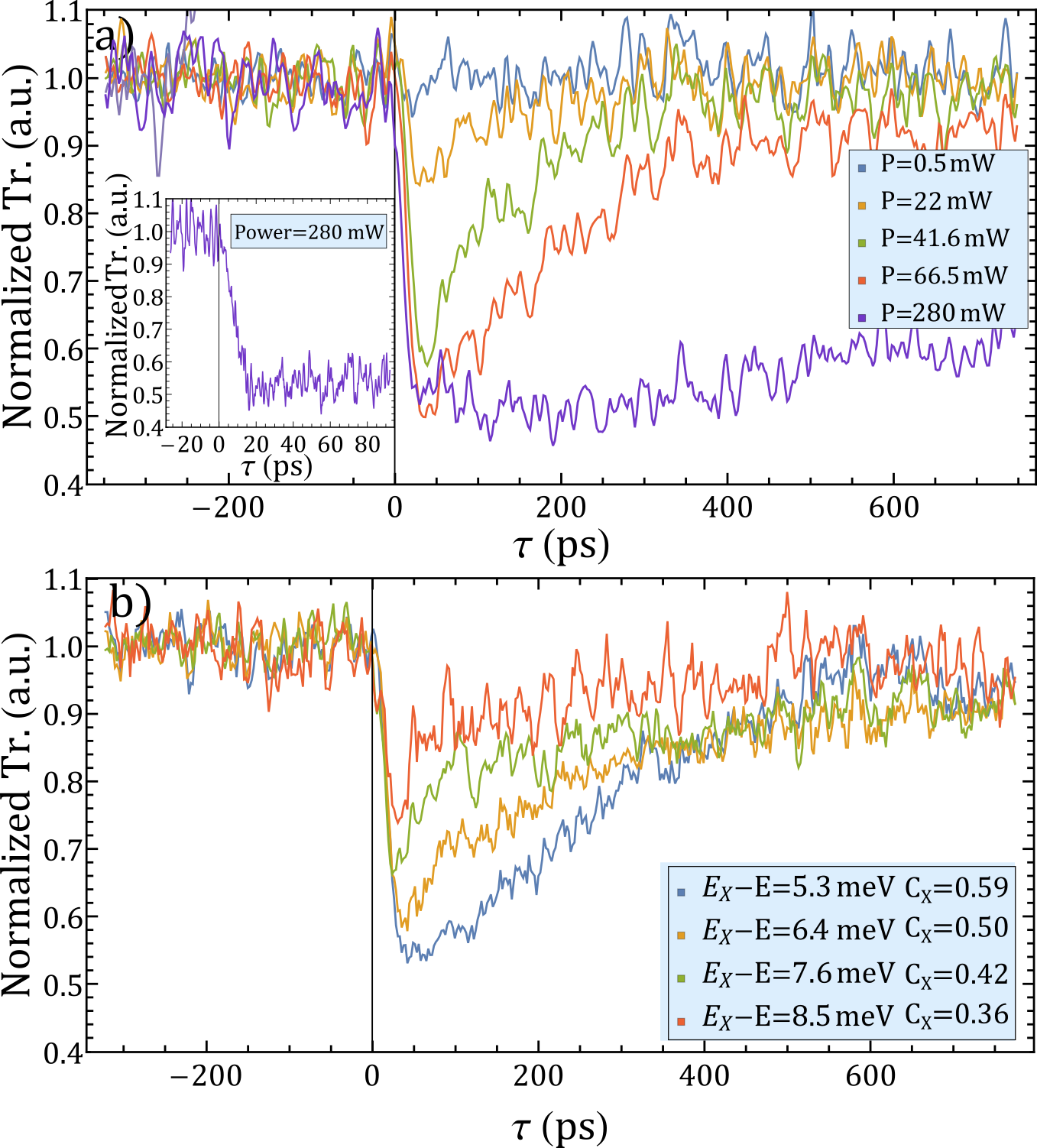}
    \caption{a) Time resolved transmission of the CW laser when injected resonantly into the WG. At $\tau=0$ the fs pulse reaches the system, modifying the excitonic resonance and hence, reducing the transmission of the CW laser by a factor $\approx2$ in less than 20 ps. The effect has a strong dependence on the fs laser power. 5 different powers are displayed in the figure. The inset shows the dynamics for a pump power of 280 mW, but measured with a higher temporal resolution, allowing to appreciate the switch-off speed. b) Analysis of the switching effect at different energies of the CW laser. The energies closer to the excitonic resonance are more affected than the photonic ones, evidencing the excitonic origin of the effect.}
    \label{trans}
\end{figure}

We start by using a CW laser as signal. The transmitted light is directed into a streak camera and we observe the modification in the transmission of the signal as a function of time for different powers of the control beam and for different energies of the signal beam. In each case the transmission is normalized to the intensity of the continuous signal during the first 200 ps ($\sim -400$ ps to $-200$ ps). The need for this normalization comes from the fact that the total CW signal at the initial time depends on the pulse power, suggesting that the system does not fully empty between consecutive pulses (12 ns), as we will discuss later (see also SM). The results are shown in Fig.~\ref{trans}. Panel a corresponds to the case in which the CW laser has an energy of 1525.14 meV; 4.9 meV below the excitonic resonance. As the fs laser power increases, the effect becomes evident: at $\tau=0$, when the pulse reaches the system, the transmission is reduced to values lower than 0.5 in less than 20 ps (see inset). 
Notice from Fig.~\ref{trans}a how a power of 66.5 mW in the pulsed laser corresponds to a switching time of less than 20 ps with a switching energy of $\approx 0.8$ nJ. The increasing recovery time with pump power, indicates the creation of a population of charge carriers through two-photons absorption. \\

The excitonic nature of the effect, implies a dependence on the energy of the continuous flux of polaritons (respect to the excitonic resonance). In the case of Fig.~\ref{trans}a, the CW population is injected at an excitonic fraction $C_X=0.69$. After fixing the pulsed laser power to 350 mW, the experiment is repeated for other excitonic fractions (Fig.~\ref{trans}b). These results confirm that the switching-off has an excitonic origin: the higher the excitonic fraction of the CW beam, the lower the minimum transmission reached after the pulse arrival.\\

The data collected with the streak camera allow to clearly reconstruct the recovery of the cw transmission after the fs control has passed. To reconstruct the much faster switch off of the transmission we use a homodyne detection technique. The fs control is used to generate a fs signal through the super-continuum (SC) generation effect (Fig.~S1). This technique allows us to achieve sub-ps temporal resolution. We reconstruct the dynamics by using a delay line that changes the arrival time of the fs control pulse respect to the probe signal (see also SM). In this regime, the Optical Stark effect governs the dynamics \cite{Hayat2012,Lange2018}. It modulates the light-matter coupling during the pulse temporal width. A delay line provides the temporal control needed to obtain a precision on the order of tens of fs. By varying the arrival time of the fs control pulse, and using the SC pulse to probe the system, we reconstruct these ultrafast dynamics ($\tau\equiv t_{SC}-t_{fs}$).
\begin{figure}
    \centering
    \includegraphics[width=0.99\columnwidth]{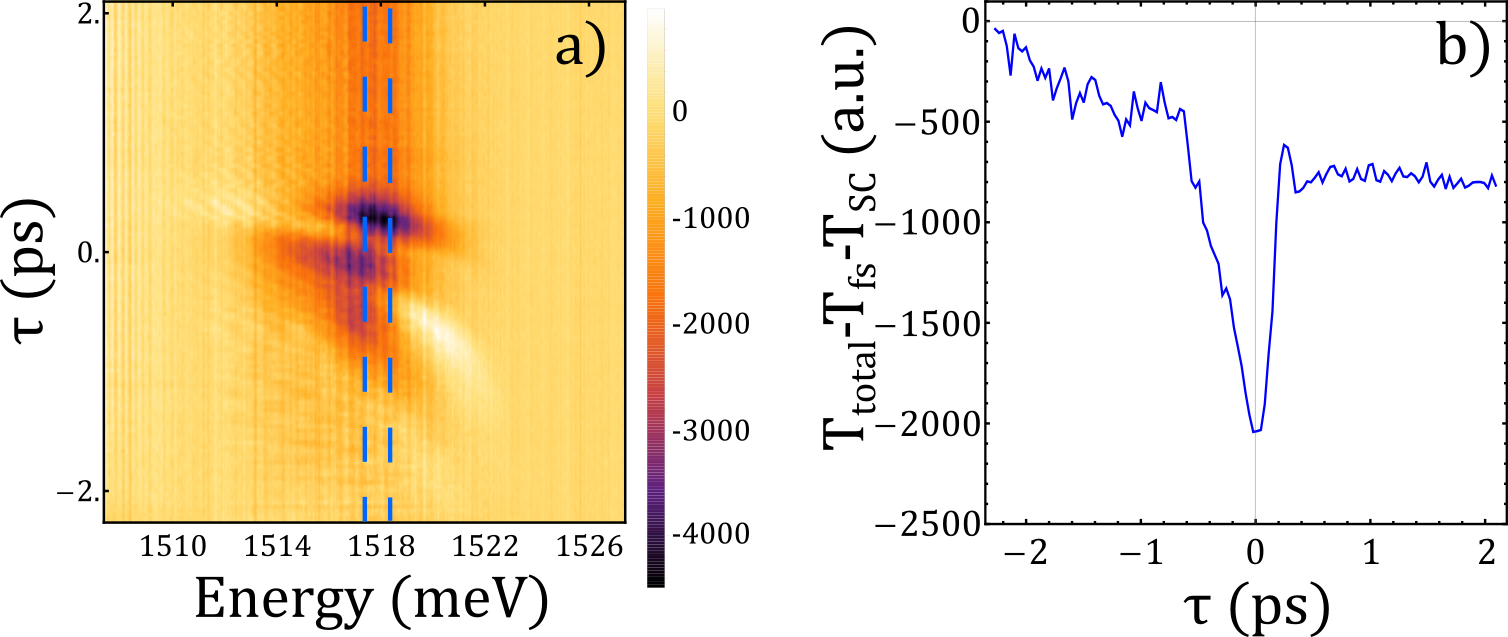}
    \caption{Characterization of the optical Stark effect in the microwire structure. a) Differential transmission of the SC laser at a fixed in-plane momentum and for different arrival times of the fs pulse. b) vertical profile at a fixed energies (dashed rectangle in panel a). With the arrival of the control pulse, the light-matter coupling is modulated, generating an instantaneous energy shift, that causes a reduction of the transmitted intensity.}
    \label{stark}
\end{figure}
The results, presented in Fig.~\ref{stark} show the differential transmission, defined as: $T_{total}-T_{fs}-T_{SC}$, where $T_{total}$ is the transmission of the SC pulse in presence of the fs pulse (for different time delays), $T_{fs}$ is the system transmission when only the fs pulse is present and $T_{SC}$ is the transmission when only the SC pulse reaches the WG. The instantaneous shifting at $\tau\sim0$ (Fig.~\ref{stark}a) is the clear signature of the optical Stark effect, that modulates the light-matter coupling, shifting the lower polariton branch. 
Panel b displays a profile taken at the original polariton energy for a given k, showing the instantaneous modification of the transmission thanks to the Stark modulation at $\tau=0$. 
For longer times, the control beam arrives before the SC pulse, generating a bath of charge carriers via two-photon absorption. This explains the lower differential transmission at $\tau>0$. Although a streak camera cannot resolve the Stark ultrafast dynamics, it allows to observe the effect of the reservoir on the transmission of a continuum polariton flux.\\

It is worth to mention that the additional transverse confinement provided by the etched microwire, is essential to observe the switching effect. The experiment was repeated in identical conditions in wider structures of 20 $\mu$m, but no reduction in the transmission was observed after the injection of the fs pulse, indicating that not only the pulse must be injected into the structure, but its modal volume has to be reduced. Moreover, by pumping with the same energy of the fs laser, but with an in-plane momentum that does not match the guided mode (and hence it does not couple into the WG), the continuous polariton flux is not interrupted after the pulse arrival, indicating that the effect is not present at the grating coupler for the two lasers. In a similar way, by tuning the energy of the pulsed laser out of the resonant condition to lower energies, the flux is not interrupted either. The population of charge carriers could also be created by tuning the fs pulse at energies above the material's band-gap. The dynamics obtained in this configuration (Fig.~S2) show the desired switching effect, but the unavoidable PL signal constitutes an undesirable noise in the detection system, hindering a practical application of this effect. Finally, no dependence on the CW power was detected.\\

As mentioned above, the initial transmission for each fs pulse power varies (see Fig.~S3), which suggests that the cavity does not fully empty between consecutive pulses. This indicates that the reservoir has an important component of dark-long-lived excitons whose lifetime is, at least, comparable with the delay between fs pulses (12 ns). Importantly, this particular behavior does not affect whatsoever the switching condition, since the important quantity in this case is the relative transmission, moreover it could be exploited to constrain the system to remain in its ``off" state during very long times compared to the switching time. This point is illustrated and discussed in the Supplementary Material. For a theoretical modelization of this effect we rely on a Gross-Pitaevskii (GP) formalism for a population of polaritons in presence of a reservoir of excitons and charge carriers\cite{Gargoubi2016}. As demonstrated in previous works, the nonlinearity has an apparent increase due to the presence of such particles\cite{Walker}, which is the origin of the energy shift that affects the polariton transmission through the WG. The 1D confinement provided by the $1$ $\mu$m wire, justifies the one-dimensional model used here. In this conditions the system behavior can be described by the set of equations \ref{GPE}.\\

\begin{widetext}
\begin{subequations}
\begin{eqnarray}
i\hbar\frac{\partial \Psi(x,t)}{\partial t}&=&\left[-\frac{\hbar^2 }{2m}\frac{\partial^2}{\partial x^2}+g_{XX}\cdot C_X\left[|\Psi(x,t)|^2+n_R(x,t)\right]-i\left(C_X\gamma_e+\left[1-C_X\right]\gamma_p\right)\right]\Psi(x,t)+P_{CW}(x,t)\\
\hbar\frac{\partial n_R(x,t)}{\partial t}&=&-2\gamma_R n_R(x,t)+P_{fs}(x,t)
\end{eqnarray}\label{GPE}
\end{subequations}
\end{widetext}

In this set of coupled equations $\Psi(x,t)$ represents the wavefunction of the polariton injected resonantly by the CW laser, $g_{XX}$ is the strengh of the exciton exchange interaction, $C_X$ the excitonic fraction, $\gamma_e$, $\gamma_p$ and $\gamma_R$ the decay rates of the (coupled) exciton, photonic mode and long-lived reservoir population, respectively. Finally, $n_R(x,t)$ represents the particles density of the reservoir. The pumping terms $P_{CW}$ and $P_{fs}$ stand for the external injection of resonant polaritonic population and exciton reservoir by two photon absorption. Their functional form will be determined by the experimental pump configuration. In our case we focus the CW (that resonantly injects population of polaritons) and fs (that indirectly injects population into the long-lived charge-carriers reservoir) lasers into the input grating at $x=0$, so (in the rotating frame) their functional form is:
\small{
\begin{subequations}
\begin{eqnarray}
P_{CW}(x,t)&=&p_{CW}e^{-\left(x/\sigma_{CW}\right)^2}e^{i\Delta_{\omega}t}\\\label{pcw}
P_{fs}(x,t)&=&N(1-e^{-\epsilon\cdot p_{fs}^2})e^{-(x/\sigma_{fs})^2}e^{-\left(\frac{t-t_o}{\tau}\right)^2}\label{pfs}
\end{eqnarray}
\end{subequations}}

\normalsize{
Where $p_{CW}$ and $p_{fs}$ are proportional to the continuum and pulsed lasers intensity, respectively, $\sigma_{CW}$ and $\sigma_{fs}$ determine the waist size of the CW and pulsed lasers, respectively, $\hbar \Delta_{\omega}$ is the energy difference between the excitonic resonance and guided mode at the in-plane momentum of the resonant pump, $t_o$ is the arrival time of the fs pulse and $\tau$ depends on its temporal width. The dependence of $P_{fs}$ with $p_{fs}$ takes into account the saturable nature of an excitonic reservoir, while the quadratic dependence of $P_{fs}$ takes into account the effect that creates the reservoir: two photons absorption with an efficiency $\epsilon$. The complete explanation of the model and a table with the values of the parameters employed in the simulations (and their experimental deduction) can be found in the SM.}\\

\begin{figure}
    \centering
    \includegraphics[width=0.85\columnwidth]{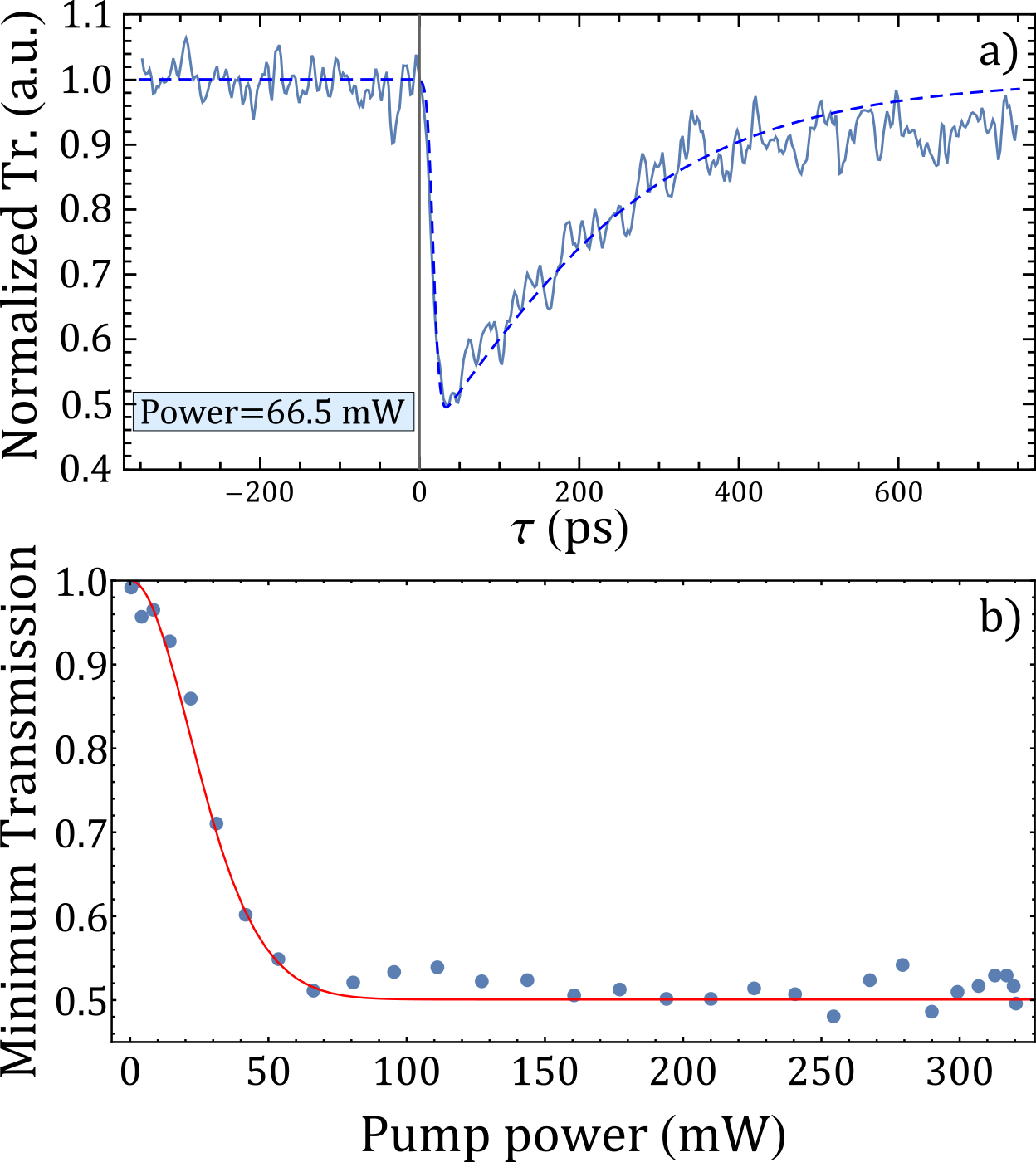}
    \caption{a) Experimental time-resolved transmission for the case of a pulsed laser power of 66.5 mW (continuous line) with the theoretical fitting from the GPE system of equations 1 (dashed line). b) Power dependence of the CW laser transmission minimum. 
    The continuous red line corresponds to the reproduction of this behavior by the theoretical model.}
    \label{minima}
\end{figure}

\normalsize{
The transmission calculated with this method and the experimental data of the time-resolved transmission for the pump power 66.5 mW are presented in Fig.~\ref{minima}a. The good correspondence between theory and experiment confirms the hyphotesis of the reduction in the transmission due to the renormalization of the energy dispersion induced by the charge-carriers reservoir.}\\ 

\normalsize{
In summary, we demonstrated an ultrafast all-optical switching of a CW laser in a polariton WG system. The ultrafast switching behaviour comprises two distinct physical effects spanning different temporal scales and powers: an ultrafast modulation of the light-matter coupling that occurs in a time range of hundreds of fs due to the optical Stark effect, and a renormalization of the polariton energy due to the creation of a charge carriers reservoir. While the former effect takes place within the duration of the pulse beam, the latter effect occurs in a time range of tens of ps. A comparison of our system with state-of-the-art all-optical switches\cite{Rutckaia2020} reveals that the system studied here is among the fastest all-optical switches demonstrated so far. Both switching effects require energies of approximately 0.8 nJ, a quantity that can be reduced by either further reducing the photonic modal volume or by the optimization of the pulse injection through the grating. These results provide  an additional step in the road towards fully optical technologies, capable to operate without the need of an optical-electronic-optical conversion.\\}

\noindent \textbf{Acknowledgments}\\
We thank Paolo Cazzato for technical support.\\
We are grateful to Prof. Ronen Rapaport for inspiring discussions and for sharing information about the sample design.\\
The authors acknowledge the Italian Ministry of University (MUR) for funding through the PRIN project ``Interacting Photons in Polariton Circuits'' – INPhoPOL (2017P9FJBS\_001).\\
Work at the Molecular Foundry was supported by the Office of Science, Office of Basic Energy Sciences, of the U.S. Department of Energy under Contract No. DE-AC02-05CH11231.\\
We thank Scott Dhuey at the Molecular Foundry for assistance with the electron beam lithography.\\
We acknowledge the project FISR - C.N.R. “Tecnopolo di nanotecnologia e fotonica per la medicina di precisione” - CUP B83B17000010001 and "Progetto Tecnopolo per la Medicina di precisione, Deliberazione della Giunta Regionale n. 2117 del 21/11/2018.\\
This research is partly funded by the Gordon and Betty Moore Foundation’s EPiQS Initiative, Grant GBMF9615 to L. N. Pfeiffer, and by the National Science Foundation MRSEC grant DMR 1420541.\\

\noindent \textbf{Competing Interest}\\
The authors declare no competing interest.

\bibliographystyle{unsrt}
\bibliography{biblio.bib}

\clearpage
\onecolumngrid

\setcounter{equation}{0}
\setcounter{figure}{0}
\setcounter{table}{0}
\setcounter{page}{1}
\makeatletter
\renewcommand{\theequation}{S\arabic{equation}}
\renewcommand{\thefigure}{S\arabic{figure}}
\pagenumbering{roman}

\noindent
\textbf{\Large Supporting Information}\\
\section{Experimental setup}
\begin{figure}[!htb]
    \centering
    \includegraphics[width=0.7\columnwidth]{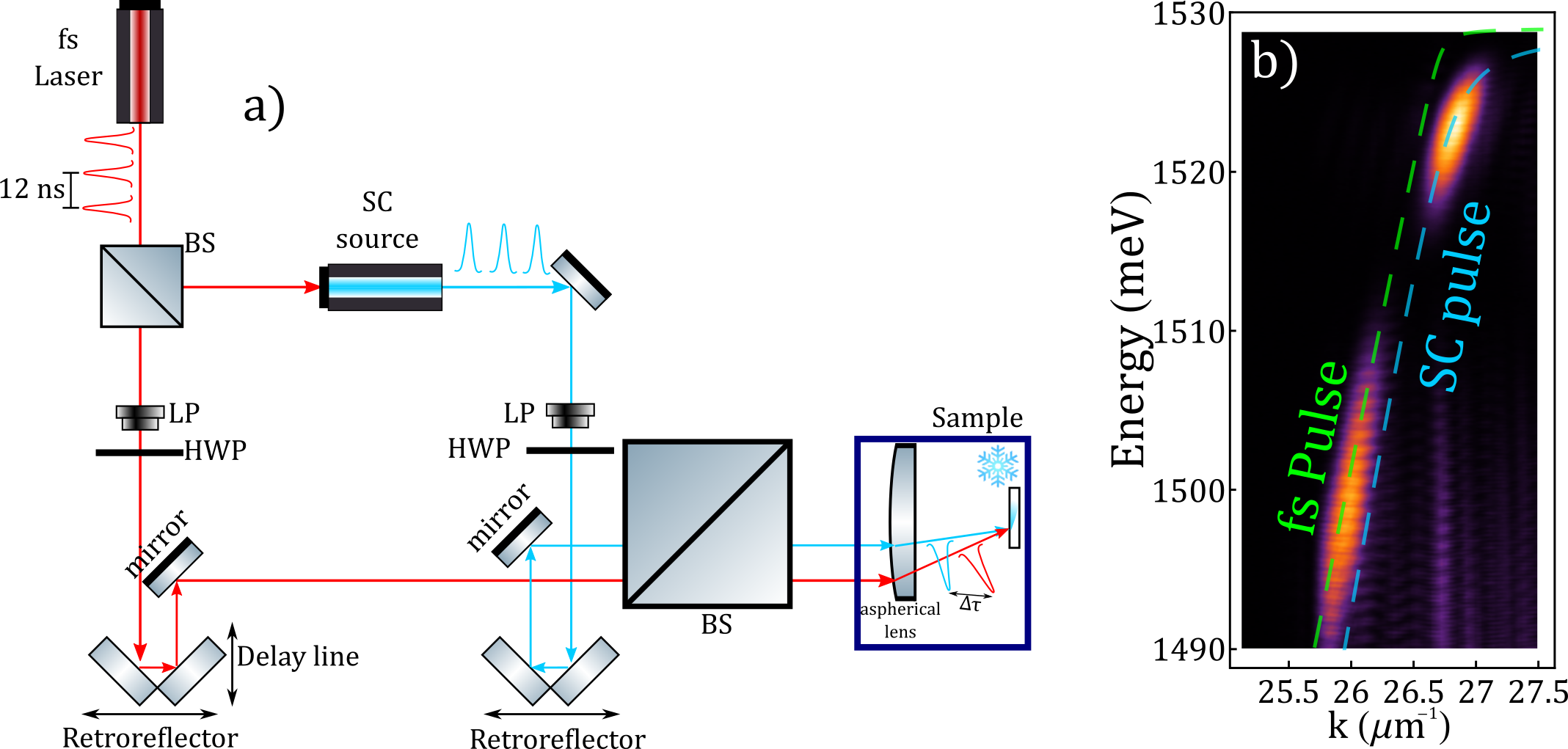}
    \caption{Characterization of the optical Stark effect. a) scheme of the pumping setup. The fs pulse feeds a supercontinuum source, generating another pulse with a broad spectrum. The desired energy is selected afterwards with a bandpass filter, and then injected resonantly into the system. b) System transmission when both pulses are injected resonantly. The control pulse is resonant with the $TM_o$ mode, while the SC laser is resonant with the $TE_o$ mode. }
    \label{stark_setup}
\end{figure}

\begin{figure}[!h]
    \centering
    \includegraphics[width=0.7\columnwidth]{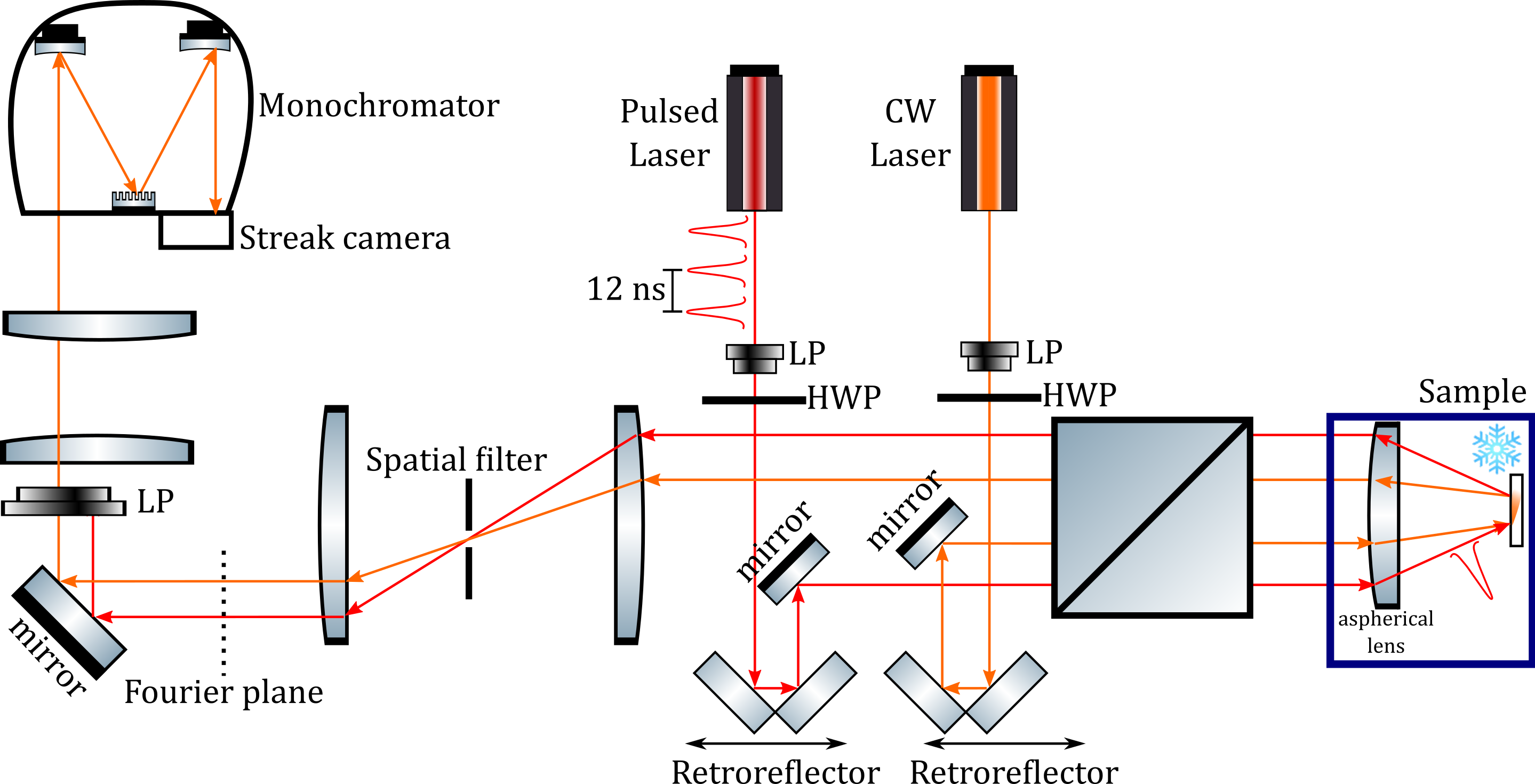}
    \caption{Depiction of the experimental setup. LP stands for "Linear Polarizer" and "HWP" stands for Half-wave plate.}
    \label{setup}
\end{figure}

Two different experimental configurations are used along this work, depending on the studied effect. The sub-picosecond modulation of the energy dispersion by the Optical Stark effect cannot be resolved with a streak camera, due to its typical time range, which depends on the pump pulse temporal extension. In our case, we use a laser with a pulse duration of 100 fs, a time window below the camera's resolution ($>$250 fs). Therefore, to characterize this effect in our system we used an ultrafast pulse, generated with a super-continuum (SC) source to probe the polariton dispersion. This configuration is illustrated in Fig.~\ref{stark_setup}a. A delay line was implemented for the fs pulse, in order to vary its arrival time respect to the SC pulse. While the SC pulse was coupled resonantly into the TE mode, the fs pulse is injected resonantly into the TM mode. The transmission of the system when both lasers are injected is displayed in Fig.~\ref{stark_setup}b.  The effect was then characterized by registering the transmission of the SC pulse as the arrival time of the fs pulse was systematically modified.\\
The picosecond regime, product of the creation of a charge-carriers reservoir via two photons absorption is displayed in Fig.~\ref{setup}. We use a tunable Ti:sapphire CW laser. For the different configurations, the laser is set at variable energies in the range of the NIR, from 1525.14 meV to 1521.55 meV. Since the aim was to pump resonantly the $TE_o$ polariton branch, after the linear polarizer the HWP rotates the polarization to match such guided mode. The pulsed laser is tuned as far to the exciton as the numerical aperture of the pumping setup (including the patterned gratings) allows. It is a Ti:sapphire with 80 MHz of repetition rate and 100 fs of pulse duration. It is tuned (centered) at around 1500 meV. Both lasers have independent optical lines, in order to have fine control on the injection angle and polarization of each one. The sample is kept at a cryogenic temperature of 4K during the full experiment, and to avoid any thermal effect (specially due to the CW pump), we use a shutter that opens 10 ms every 3 seconds and we trigger the streak camera to open in synchrony.\\
After the injection, propagation and extraction of the optical signals, the real space plane is reconstructed and the reflection of the input lasers is suppressed with a slit, so only the light from the outcoupling grating is collected in detection. After the real space filter, the Fourier plane is focused on a monochromator+streak camera system. The signal from the pulsed laser is focused on the detection system to synchronize the time of arrival of the pulse, after which it is filtered in energy (by the monochromator), polarization (with the linear polarizer) and angle (since CW and pulsed signals are resonant with the guided modes at different in-plane momenta, they are injected and extracted with different angle). The CW signal is then analyzed in the streak camera with a temporal resolution that varies from 0.25 ps to 2.3 ps, depending on the selected time range.

\section{Pulse energy above the bandgap}
\begin{figure}
    \centering
    \includegraphics[width=0.5\columnwidth]{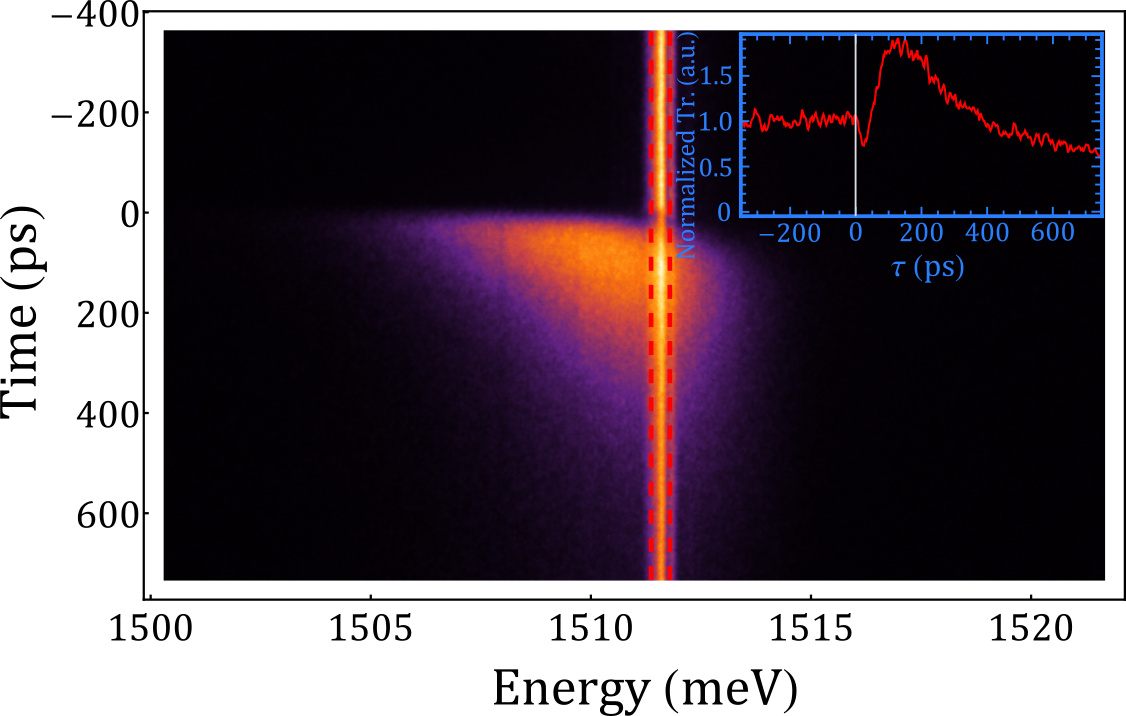}
    \caption{Time resolved output signal for the case in which the fs pulsed laser is tuned at an energy above the QW bandgap and with a power of $100$ mW. The PL emission composes a great part of the detected light, adding undesired noise to the CW signal. The inset corresponds to the energy selection indicated by the red dashed lines, which is the CW signal affected by the light generated by PL emission.}
    \label{switch_pl}
\end{figure}

As mentioned in the main text, the switch effect has its origin in an instantaneous modification of the polaritonic dispersion, which takes the coupled CW signal out of resonance, reducing its transmission. Since such modification is due to the creation of a charge carrier reservoir, one might think that it is more efficient to generate directly the reservoir by pumping with a pulse whose energy overcomes the QW bandgap. This indeed creates a larger reservoir, however, it is not the best option, and the reason can be understood by looking at Fig.~\ref{switch_pl}.
The fs pulse generates a PL emission that overlaps with the CW signal, adding considering noise to the output. There is still a reminiscence of the switching effect (as displayed in the inset), but it is critically affected by the added optical noise. Although this configuration does not represent the best option for the implementation of the ultrafast switch of the CW, it is in good agreement with a transmission reduction originated by an excitonic shift and consequent polariton dispersion renormalization.

\section{Formation and effects of Dark Excitons}
\begin{figure}
    \centering
    \includegraphics[width=0.6\columnwidth]{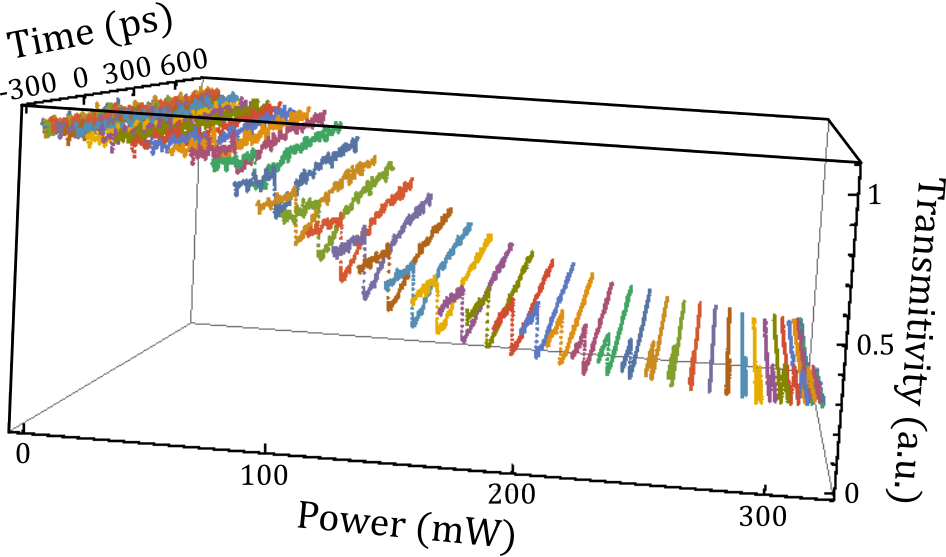}
    \caption{Time-resolved unnormalized transmission of CW laser for different powers of the pulsed laser. The reduction of the initial transmitivity for increasing powers witnesses the creation of a reservoir of long-lived dark excitons.}
    \label{DE}
\end{figure}

As it has been previously discussed, the origin of the switching effect is the creation of a charge reservoir by two-photons absorption of the pulse after injected inside the structure. Such reservoir is composed, among others, by dark excitons, whose optical recombination is prevented by the spin selection rules. The effects of such particles in the optical properties of the material is rather critical, though. As mentioned in the main text, a large population of dark excitons could carry a renormalization of the polaritonic dispersion, very hard to quantify due to the lack of emission from these particles. They are characterized by very long lifetimes \cite{Kavokin2008}. In our case, after a critical power of the pulsed laser, such states become relevant, and affect the transmission in time ranges longer than the delay between pulses. The initial transmission of the CW laser is then affected by the presence of dark-excitons accumulated in consecutive pulses. It is worth to mention that the ``switching" condition is not affected whatsoever by this effect, since, independent from the initial transmission, the CW laser is reduced up to half its initial intensity. The reservoir of dark excitons implies then a renormalization of the CW flux transmission, changing, precisely, the CW flux inital condition. The details of such behavior are displayed in Fig.~\ref{DE}.\\

Notice how the initial condition of the CW laser is modified by the pumping power but the switching events still takes place. This suggests that the reservoir does not totally empty between consecutive pulses, indicating that the created dark excitons have lifetimes at least comparable with the delay between laser pulses (12 ns).

\section{Theoretical model}
For the theoretical modelization of the polariton scattering by the excitonic reservoir we implement a Gross-Pitaevskii model for a density of polaritons in presence of a reservoir of long-lived, uncoupled excitons. The system of equations describing the system is as follows\cite{Gargoubi2016}:

\begin{subequations}
\begin{eqnarray}
i\hbar\frac{\partial \Psi(x,t)}{\partial t}&=&\left[-\frac{\hbar^2 }{2m}\frac{\partial^2}{\partial x^2}+g_{XX}\cdot C_X\left[|\Psi(x,t)|^2+n_R(x,t)\right]-i\left(C_X\gamma_e+\left[1-C_X\right]\gamma_p\right)\right]\Psi(x,t)+P_{CW}(x,t)\label{gpea}\\
\hbar\frac{\partial n_R(x,t)}{\partial t}&=&-2\gamma_R n_R(x,t)+P_{fs}(x,t)\label{gpeb}
\end{eqnarray}\label{GPE_s}
\end{subequations}

Where $\Psi(x,t)$ is the wavefunction of the polariton injected resonantly. The horizontal  (y) dependence can be neglected due to the additional confinement provided by the microwire, that restrains the mode to 1D. The polariton mass $m$ has tricky details. 
If it is calculated as $\frac{\partial^2 E}{\partial k^2}$ the parameter will take negative values. We decided to employ a common value for a DBR microcavity: $3\times 10^{-5}m_e$, where $m_e$ is the free electron mass. $g_{XX}$ is the strength of the exchange exciton-exciton interaction, whose value has been widely debated and very different values have been reported in different experiments\cite{Estrecho2019}. In this work we took as reference the commonly accepted value of $6$ $\mu eV\cdot \mu m$\cite{Tassone1999}. The excitonic fraction $C_X$ determines how close in energy the injected CW polaritons are from the excitonic resonance, and hence, how strong the interaction is \cite{Kavokin2008}. The decay rates of the excitonic ($\gamma_e$) and photonic ($\gamma_p$) modes are determined experimentally by pumping the WG system out of resonance while systematically vary the distance from the pumping spot to the outcoupling grating. The reduction of the PL intensity with the pumping distance, allows to determine the decay rate of each polariton component. $\gamma_R$, the decay rate of the exciton reservoir is much more challenging to determine, given the dark nature of a large portion of the composing particles, therefore, this parameter is used as a fitting degree of freedom. In equation \ref{gpeb} $n_R(x,t)$ is the reservoir particles density. The pumping terms $P_{CW}(x,t)$ and $P_{fs}(x,t)$ depend on the details of each laser. In particular they have the form:
\begin{subequations}
\begin{eqnarray}
P_{CW}(x,t)&=&p_{CW}e^{-\left(x/2\sigma_{CW}\right)^2}e^{i\Delta_{\omega}t}\\
P_{fs}(x,t)&=&N(1-e^{-\epsilon\cdot p_{fs}^2})e^{-(x/2\sigma_{fs})^2}e^{-\left(\frac{t-t_o}{\tau}\right)^2}
\end{eqnarray}
\end{subequations}

In the rotating frame, $\Delta_{\omega}$ represents the difference between the exction and the photonic mode energies at the resonant in-plane momentum. As it can be observed, the temporal dependence of each expression varies due to the nature of the pump: the CW case is represented by an harmonic function, while the pulsed laser has a temporal gaussian profile whose amplitude depends on the pulse duration. Consider that this time must be much longer than the pulse temporal width, because it takes into account the formation of the reservoir. $\sigma_{CW}$ and $\sigma_{fs}$ depend on the spatial profile of the CW and pulsed lasers, respectively. Both values can be easily determined from an imaging of the real space plane. The pump powers fix the values $p_{CW}$ and $p_{fs}$, however, their determination is not straightforward for the following reasons: on the one hand the in-coupling and out-coupling efficiency of the etched gratings cannot be precisely measured; it is not correct to assume that the coupling efficiency is equal to the extraction one, since the spatial profiles in each case are fundamentally different (gaussian for the input and exponential for the output). On the other hand, the exact amount of the pulsed energy that becomes dark excitons is difficult to obtain due to the lack of an optical signal. The pump parameters $p_{fs}$ and $p_{CW}$ are then treated as fitting parameters. As mentioned in the main text, the dynamics has a negligible dependence on $p_{CW}$, making this a minor issue. Notice the particular dependence of $P_{fs}$ with $p_{fs}$; this expression takes into account the saturability of the bath of excitons. Pauli exclusion principle prevents the formation of two excitons separated by a distance shorter than the Bohr radius\cite{Carusotto2013}, precluding the formation of a reservoir with a population higher than $N$. Since the process by which the bath of charge carriers is created is two-photons absorption, the process have a quadratic dependence with $p_{fs}$ and it has an efficiency determined by $\epsilon$. The values of the parameters employed for the numerical simulation are summarized in Table \ref{tab}.

\begin{table}[!h]
\begin{tabular}{|c|c|c|c|c|c|c|c|c|c|c|c|}\hline
$m$&$g_{XX}$&$C_{X}$&$\gamma_e$&$\gamma_p$&$\gamma_R$&$\sigma_{CW}$&$\sigma{fs}$&$p_{CW}$&$N$&$\epsilon$&$\tau$\\ \hline
$3\cdot 10^{-5}m_e$&$6$ $\mu eV\cdot\mu m$&$0.69$&$223$ $\mu eV$& $30$ $\mu eV$&$1.8$ $\mu eV$&$35$ $\mu m$&$35$ $\mu m$&$0.3$ $eV\cdot\mu m^{-1/2}$&$1.8$ $eV\cdot\mu m^{-1}$&$7\cdot 10^{-4}mW^{-2}$&$80$ ps\\ \hline

\end{tabular}\caption{Parameters employed in the numerical simulations.}\label{tab}
\end{table}

\end{document}